\documentclass[12pt,preprint]{aastex}

\usepackage{color}
\usepackage{amsmath}

\shorttitle{Analysis of the breakup of P/2013 R3}
\shortauthors{Hirabayashi et al.}

\begin{document}


\title{Constraints on the Physical Properties of Main Belt Comet P/2013 R3 from its Breakup Event}


\author{Masatoshi Hirabayashi\altaffilmark{1}}
\affil{Aerospace Engineering Sciences, 429 UCB, University of Colorado, Boulder, CO 80309-5004 United States}
\email{masatoshi.hirabayashi@colorado.edu}

\author{Daniel J. Scheeres\altaffilmark{2}}
\affil{Aerospace Engineering Sciences, 429 UCB, University of Colorado, Boulder, CO 80309-0429 United States}

\and

\author{Diego Paul S\'anchez\altaffilmark{3} and Travis Gabriel\altaffilmark{4}}
\affil{Aerospace Engineering Sciences, 429 UCB, University of Colorado, Boulder, CO 80309-5004 United States}


\altaffiltext{1}{Ph.D. candidate, Colorado Center for Astrodynamics Research, 
Aerospace Engineering Sciences, University of Colorado Boulder}

\altaffiltext{2}{A. Richard Seebass Chair, Professor, Colorado Center for Astrodynamics Research, 
Aerospace Engineering Sciences, University of Colorado Boulder}

\altaffiltext{3}{Research Associate, Colorado Center for Astrodynamics Research, 
Aerospace Engineering Sciences, University of Colorado Boulder}

\altaffiltext{4}{Ph.D. student, Colorado Center for Astrodynamics Research, 
Aerospace Engineering Sciences, University of Colorado Boulder}


\begin{abstract}
\cite{Jewitt2014B} recently reported that main belt comet P/2013 R3 experienced a breakup, probably due to rotational disruption, with its components separating on mutually hyperbolic orbits. We propose a technique for constraining physical properties of the proto-body, especially the initial spin period and cohesive strength, as a function of the body's estimated size and density. The breakup conditions are developed by combining mutual orbit dynamics of the smaller components and the failure condition of the proto-body. Given a proto-body with a bulk density ranging from 1000 kg/m$^3$ to 1500 kg/m$^3$ (a typical range of the bulk density of C-type asteroids), we obtain possible values of the cohesive strength (40 - 210 Pa) and the initial spin state (0.48 - 1.9 hr). From this result, we conclude that although the proto-body could have been a rubble pile, it was likely spinning beyond its gravitational binding limit and would have needed cohesive strength to hold itself together. Additional observations of P/2013 R3 will enable stronger constraints on this event, and the present technique will be able to give more precise estimates of its internal structure.
\end{abstract}





\section{Introduction}
Recent years have seen several observations of a previously unrecognized class of small bodies, what have been called ``Active Asteroids.'' These are bodies that display intermittent phenomenon traditionally associated with comets, yet which appear for bodies that are thought to be asteroids with no or minimal volatiles. A survey of recent observations is given in \cite{Jewitt2012}; however, since that time, there have been several additional and striking examples of this phenomenon \citep{Jewitt2013, Jewitt2014B}. It is important to note that the observed characteristics of these active asteroids are not uniformly similar. An excellent example of this is the contrast between bodies P/2013 P5 and P/2013 R3. The former was observed to have several streamers of dusty material emanating from a single main body at several different epochs in relatively close spacing. The latter, however, was observed to be components that were mutually escaping from each other, with these individual components undergoing additional fractures at later epochs. While the root cause of these events is thought to be rotational disruption \citep{Jewitt2013, Jewitt2014B}, the differing observed morphologies may indicate different modes of failure (e.g., \cite{Hirabayashi2014}). Specifically, while P/2013 P5 may be indicative of mass-shedding of regolith from the surface of an asteroid, P/2013 R3 appears to be consistent with a body breaking into multiple components. In this paper we focus on this latter active asteroid and, under the hypothesis that it was a single asteroid that underwent rotational disruption, develop constraints on its physical properties. 

Analytical modeling of this body can provide clues about the origin and mechanism of these events. We explore a model for the breakup of an ellipsoidal rubble pile that was firstly discussed in \cite{Scheeres2010} and that was later expanded in \cite{Sanchez2014}. In the present model a body that has cohesion can be spun beyond the rate at which centrifugal accelerations can be balanced with mutual gravitational attractions. Depending on the strength of the cohesive bonds, which may be less than a few hundred pascals \citep{Sanchez2014}, at some spin rate the body may fracture along planes of weakness, with the components then departing each other on possibly hyperbolic orbits. The trigger for the fracture may be either a secular increase in spin rate due to the YORP effect, or a small impact that generates seismic waves that cause bonds close to the failure limit to fail \citep{Marzari2011}. 

Four images of P/2013 R3 taken at different epochs between October and December in 2013 show that this object experienced subsequent breakups \citep{Jewitt2014B}. It is reported that the proto-body has broken into more than 10 components as a result of this breakup event. The maximum size of the components may be on the order of a few hundred meters. In this paper we assume that the YORP effect causes the body to spin up to its critical spin, so detailed discussions of the spin-up mechanism are omitted. Based on the observational estimates by \cite{Jewitt2014B}, we derive the initial spin period and the cohesive strength of P/2013 R3. Additional observations of this system and more precise astrometric analysis of the observations may provide further constraints on the body. 

\section{Modeling of the Breakup Process}
\subsection{Breakup Scenario}
Suppose that the proto-body uniformly rotates in a principal axis mode and is only subject to its self-gravitational, centrifugal, frictional and cohesive forces. If the proto-body spins fast enough, it can fail structurally and break up into multiple components. The proto-body and the smaller components as a result of the breakup are assumed to be a biaxial ellipsoid and spheres, respectively. If shear strength is zero over some cross section, the breakup should start from this cross section. However, if shear strength is nonzero, the body will keep its shape at a faster spin rate. Applying the Mohr-Coulomb yield criterion (e.g., \cite{Chen1988}), we represent shear strength by a friction angle and cohesive strength. After a breakup, the components as a result of this event are inserted into their mutual orbits. 

\begin{figure}[hb]
  \centering
  \includegraphics[width=4in]{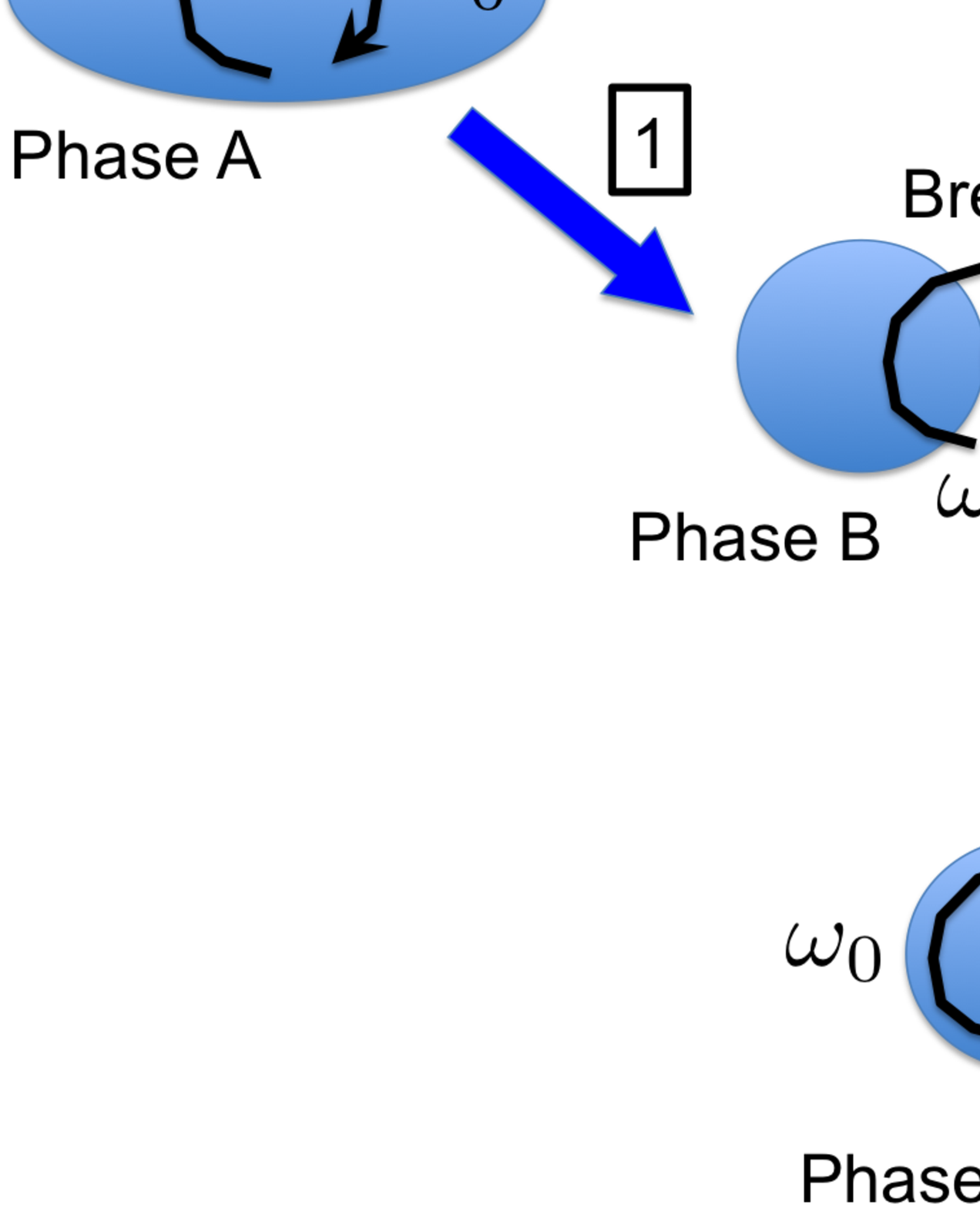}
  \caption{A model for a breakup. The proto-body (phase A) would break into two components (phase B) at the critical spin period, followed by orbital motions (phase C). At phase C, the different components may also be split, but would have mutual speeds that are lower. This event consists of two processes: process 1 being the transition from phase A to phase B and process 2 being that from phase B to phase C.}
  \label{Fig:breakup}
\end{figure}

The breakup model shown in Fig. \ref{Fig:breakup} defines two processes. Process 1 represents mechanical failure of the proto-body. The condition of this failure mode will be determined by considering the yield condition of the averaged stress over the central cross section. Process 2 describes its subsequent orbital motion as a result of mechanical failure. In general, since each component may be non-spherical, there is angular momentum transfer, resulting in a change of the spin vector during the initial disruption \citep{Scheeres2000}; however, the critical region for such a transfer of angular momentum is only at distances of a few radii of each component \citep{Hirabayashi2013}. For the case of P/2013 R3, since the initial velocity of the components are consistent with the system escape velocity \citep{Jewitt2014B}, the shape effect on the transfer is negligible. Therefore, it is a reasonable simplification of the model to assume that each component is a sphere and that the initial spin state is conserved across the breakup. 

We also assume that the entire process occurs in a plane. The initial condition of the translational velocity is determined by multiplying the spin rate by the relative distance between the centers of mass of the two components about to split, and its direction is taken to be perpendicular to the angular velocity of the proto-body. While there is evidence that fast rotators may tumble \citep{Pravec2005}, since the normal vector of a failing cross section is always on the orbital plane, the critical component of the spin vector is identical to the component normal to the orbital plane; therefore, calculations of cohesive strength should be independent of other spin components and the initial spin rate given here is equivalent to its lower bound.

It is emphasized that the terms ``breakup" and ``structural failure", used by \cite{Hirabayashi2014}, are distinguished in this study. The term ``breakup" describes that a proto-body is split into smaller pieces, while the term ``structural failure" indicates structural instability, meaning that the original shape permanently changes due to large plastic deformation, but does not necessarily break up into multiple components. If a body experiences centrifugal accelerations exceeding gravitational accelerations, any deformation may lead to break up. A single breakup is discussed here, although the present technique can be applicable to any similar cascade of breakup events. We note that the dispersion velocity between different components is proportional to their relative distance. Thus, late-separate components may have split at a similar time as the main component; however, since their speeds are less, they may not have been distinguished until later. 

We suppose that the dimensions of the proto-body are $2 a$ by $2 a \beta$ by $2 a \beta$, where $0 < \beta < 1$, and the diameters of the smaller components, denoted as $R$, are chosen to be equal to a half of the volume of the proto-body. The proto-body uniformly rotates with a spin rate $\omega$ along its maximum principal axis. The density, $\rho$, is constant over the body. 

\subsection{Structural Breakup Condition (Process 1)}
The assumption of equal sizes of the smaller components implies that the breakup occurs in the middle of the proto-body. This comes from the fact that the central cross section normal to the minimum principal axis is the most sensitive to failure. Table 2 in \cite{Jewitt2014B}, showing the effective radii of the smaller components, also indicates that the components look similar. 

To obtain the breakup condition, we use the gravitational potential at an arbitrary point inside a biaxial ellipsoid, which is written as
\begin{eqnarray}
U (x, y, z)  = \pi \rho G ( - A_0 a^2 + A_x x^2 + A_y y^2 + A_z z^2), \label{Eq:Potential}
\end{eqnarray}
where 
\begin{eqnarray}
A_0 &=& \beta^2 \int_0^\infty \frac{d s}{\Delta}, \label{Eq:A0} \\
A_x &=& \beta^2 \int_0^\infty \frac{d s}{(s+1) \Delta}, \label{Eq:Ax} \\
A_y &=& A_z = \beta^2 \int_0^\infty \frac{d s}{(s+\beta^2) \Delta}, \label{Eq:Az} 
\end{eqnarray}
and $\Delta = \sqrt{s + 1} (s+\beta^2)$. See the details in \cite{Hirabayashi2014B}.

The failure condition of the central cross section is given by considering the yield condition of the area stress over this cross section. To calculate the area stress, we use the technique proposed by \cite{Davidsson2001}. The yield condition is characterized by the Mohr-Coulomb yield criterion, which is written as 
\begin{eqnarray}
2 Y \ge (\sigma_1 - \sigma_3) \sec \phi + (\sigma_1 + \sigma_3) \tan \phi,  \label{Eq:MC}
\end{eqnarray}
where $\sigma_i \: (\sigma_1 > \sigma_2 > \sigma_3)$ is the principal components of the area stress, $\phi$ is a friction angle, and $Y$ is cohesive strength. Since the friction angle of a typical soil material ranges between 30$^\circ$ and 45$^\circ$ \citep{Lambe1969}, by taking the mean of these friction angles, the spin rate of the yield condition, $\omega_p$, is approximately described as
\begin{eqnarray}
\omega_p \sim \sqrt{\frac{4 Y}{\rho a^2} + 2 \pi \rho G A_x}, \label{Eq:omega_p}
\end{eqnarray}
where $G$ is the gravitational constant and $A_x$ is described in Eq. (\ref{Eq:Ax}). If $Y=0$, $\omega_p = \sqrt{2 \pi \rho G A_x}$, corresponding to the condition at which the pressure on the central cross section becomes zero and at which a breakup occurs. This is identical to the highest spin rate of structural failure of a cohesionless ellipsoid \citep{Sharma2009}. At this condition, the components as a result of a breakup do not fly off, but rest on each other without contact forces instead. We note that such configurations are dynamically unstable and can lead to escape after an extended period of dynamical interaction \citep{Scheeres2009, Jacobson2009, Pravec2010}. However, the case of P/2013 R3 is not consistent with this scenario as the components are seen to be immediately escaping from each other. For a body with cohesion, the spin rate of its breakup can be higher than that of any structural failure conditions and is high enough to lead to immediate escape (see the discussion on the lower size limit of binaries in \cite{Sanchez2014}). 

\subsection{Mutual Orbit After the Breakup (Process 2)}
The dispersion velocity of P/2013 R3 is 0.2 - 0.5 m/s at a relative distance of 3060 km (see Table \ref{Table:PhysicalProperty}), which is beyond the Hill sphere of the system, less than 250 km. This indicates that the small components are likely inserted in a hyperbolic orbit. Under this assumption, the following discussion explores the mutual orbit dynamics. The total energy conservation for irregular bodies is written as \citep{Scheeres2002}, 
\begin{eqnarray}
E = \int_B \frac{v^2}{2} dm - \frac{1}{2} \int_B \int_B \frac{G dm dm}{r}, \label{Eq:EnergyConservation}
\end{eqnarray}
where $B$ indicates the entire body distribution. Note that the second term in Eq. (\ref{Eq:EnergyConservation}) includes self-gravity potentials. 

Consider the initial state, i.e., the configuration where P/2013 R3 is about to break up. Assuming that the proto-body is an ellipsoid with dimensions of $a \times a \beta \times a \beta$ yields
\begin{eqnarray}
E_{initial} =  \frac{1}{2} I \omega_0^2 + \frac{\rho}{2} \int_V U(x,y,z) dV, \label{Eq:EnInt}
\end{eqnarray}
where $\omega_0$ is the initial spin rate, $I$ is the moment of inertia of the proto-body, and $U(x,y,z)$ is given in Eq. (\ref{Eq:Potential}). The first term on the right hand side in Eq. (\ref{Eq:EnInt}) is given as 
\begin{eqnarray}
 \frac{1}{2} I \omega_0^2 = \frac{1}{10} M a^2 (1+\beta^2) \omega_0^2,
\end{eqnarray} 
where $M$ indicates the mass of the whole system, i.e., $M=4 \pi \rho a^3 \beta^2/3$. The self-potential of an ellipsoid can be given as \citep{Scheeres2004}
\begin{eqnarray}
 \frac{\rho}{2} \int_V U(x,y,z) dV &=& - \frac{2}{5} M \pi \rho G a^2 A_0,
\end{eqnarray}
where $A_0$ is introduced in Eq. (\ref{Eq:A0}).  

For the configuration at a post-disruption epoch, assuming that the components are spheres with the same radius $R = a (\beta^2/2)^{1/3}$, we describe the total energy as 
\begin{eqnarray}
E_{last} &=& \frac{m_1 m_2 \Delta v^2}{2 M}  + \frac{1}{2} I_1 \omega_0^2 + \frac{1}{2} I_2 \omega_0^2 \nonumber \\
&& - \frac{G m_1 m_2}{d}  - \frac{G}{2} \int_{m_1} \int_{m_1} \frac{dm_1 dm_1}{r} -  \frac{G}{2} \int_{m_2} \int_{m_2} \frac{dm_2 dm_2}{r}, \nonumber \\
&=& \frac{1}{8} M \Delta v^2 - \frac{G M^2}{4 d} + \frac{1}{5} M R^2 \omega_0^2 - \frac{2}{5} M \pi \rho G R^2,
\end{eqnarray}
where $m_1 = m_2 = M/2$, $I_1 = I_2 = M R^2/5$, $\Delta v$ is the dispersion velocity, and $d$ is the relative distance between two components at a given epoch.

From energy conservation, $E_{initial} = E_{last}$, leading to,
\begin{eqnarray}
&& \frac{1}{5} a^2 (1+\beta^2) \omega_0^2 - \frac{4}{5} \pi \rho G a^2 A_0 \nonumber \\
&=& \frac{1}{4} \Delta v^2 - \frac{G M}{2 d} + \frac{2}{5} R^2 \omega_0^2 - \frac{4}{5} \pi \rho G R^2.
\end{eqnarray}
This relation can be solved for the initial spin rate $\omega_0$ as
\begin{eqnarray}
\omega_0 = \sqrt{\frac{\Phi}{\Psi}}, \label{Eq:ellipsoid}
\end{eqnarray}
where 
\begin{eqnarray}
\Psi &=& \frac{a^2}{5} \left( (1+\beta^2) - 2 \left( \frac{\beta^2}{2} \right)^\frac{2}{3} \right) , \nonumber \\
\Phi &=& \frac{\Delta v^2}{4} - \frac{2 \pi \rho G a^3 \beta^2}{3 d} - \frac{4 \pi \rho G a^2}{5} \left( \frac{\beta^2}{2} \right)^\frac{2}{3} + \frac{4 \pi \rho G a^2 A_0}{5}. \nonumber
\end{eqnarray}
Assuming that Eq. (\ref{Eq:omega_p}) equals Eq. (\ref{Eq:ellipsoid}), the initial spin state can also be related to the minimum level of cohesion needed for this body.  

\section{Application to P/2013 R3}
The observations by \cite{Jewitt2014B} provide the relative velocity and the distance between the components at some epochs (Table \ref{Table:PhysicalProperty}), although angle-of-view effects affect their plane-of-sky projections (Jewitt, personal communications, 2014). For the diameter of the proto-body, \cite{Jewitt2014B} reported that since they only measured the product, Area$\times$Albedo = $\pi$ Radius$^2 \times$Albedo, if the albedo is different, so is the estimated radius. The assumption of an albedo of 0.05 renders radius uncertainties by a factor of $\sqrt{2}$. Furthermore, although there is less dust in December than in October, there is still no guarantee that the dust has gone in December. Therefore, a radius from 0.2 km to 0.5 km defines the lowest and highest possibility of the radius of the proto-body\footnote{A radius of 0.2 km is equal to the radius of the largest component, while that of 0.5 km is computed from the effective radii on October 1, 2013.}. However, based on the estimates on December 13, 2013, it is strongly suspected that the effective radius may be less than 0.35 km (Jewitt 2014, personal communication). On the other hand, the aspect ratios of asteroids, $\beta$, is considered to be larger than 0.5 (c.f., the asteroid LightCurve Data Base by Warner, Harris, and Pravec, revised on November 10, 2012). Given these estimates, the present analysis provides lower bounds on the initial spin period and cohesive strength. 

Consider the first breakup event that occurred before October 1, 2013. Equations (\ref{Eq:omega_p}) and (\ref{Eq:ellipsoid}) provide the spin period of the proto-body relative to cohesive strength with different dispersion velocities and aspect ratios (Fig. \ref{Fig:initialPeriod_Cohesion}). In the figure, the solid lines show the case $\beta=0.5$, while the dashed lines describe the case $\beta=1.0$. The upper curves are the possibly slowest spin periods that result from a dispersion velocity of 0.2 m/s and a size of 1.0 km, while the lower curves are the possibly fastest spin periods that result from a dispersion velocity of 0.5 m/s and a size of 0.4 km. The actual initial spin period should be laid between the fastest and slowest spin periods. On a given curve, the bulk density increases as the spin period becomes shorter. The empty triangles and squares show the cases $\rho=1000$ kg/m$^3$ and $\rho=1500$ kg/m$^3$, respectively. If the density of this asteroid ranges between these values, then the initial spin period and cohesive strength are further constrained. From the distribution of the triangles and squares, for P/2013 R3, the cohesive strength ranges from 40 Pa to 210 Pa, while the initial spin period is between 0.48 hr and 1.9 hr. Note that in the range of friction angles of typical soils, i.e., 30$^\circ$ and 45$^\circ$, the cohesive strength changes up to a 20 $\%$ of the given values, while the initial spin period does not. 

\section{Discussion}
\begin{figure}[hb]
  \centering
  \includegraphics[width=4in]{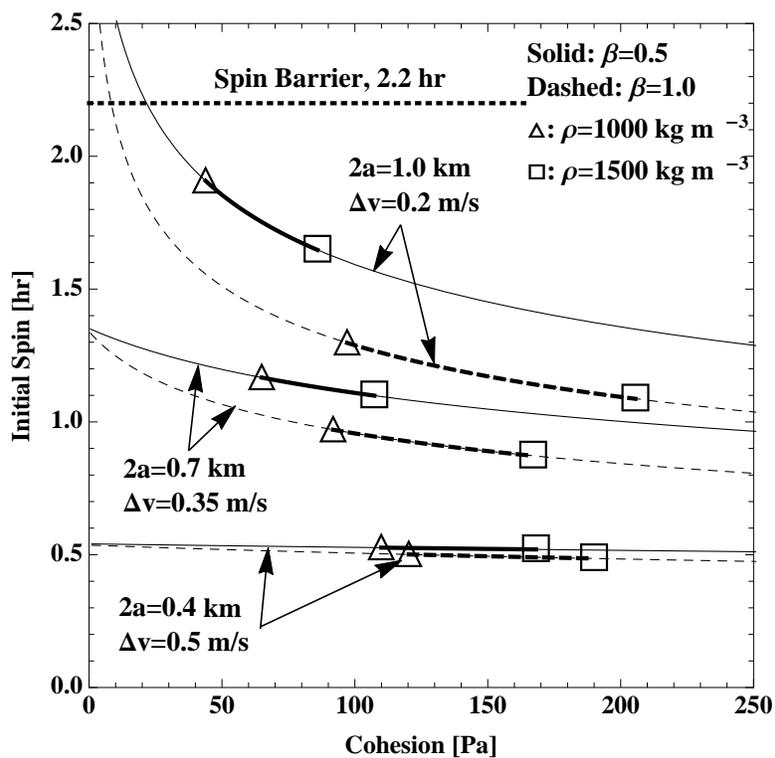}
  \caption{Possible initial spin period due to different dispersion velocities and initial sizes, i.e., $\Delta v$ ranging from $0.2$ m/s to $0.5$ m/s and $2 a$ from 0.4 km to 1.0 km (see Table \ref{Table:PhysicalProperty}). The solid lines show the initial spin period with $\beta=0.5$, while the dashed lines describe that with $\beta=1.0$. The actual spin periods should be laid between the fastest and slowest spin periods. The empty triangles and squares indicate bulk densities of 1000 kg/m$^3$ and 1500 kg/m$^3$, respectively; as a C-type asteroid, this asteroid should be between these points.}
  \label{Fig:initialPeriod_Cohesion}
\end{figure}

Broadband optical colors show that this object may be a C-type asteroid \citep{Jewitt2014B}. This type of an asteroid has relatively low bulk densities; for example, the bulk density of (206) Mathilde ranges between 1100 kg/m$^3$ and 1500 kg/m$^3$ \citep{Yeomans1997}, while that of (101955) Bennu is on the order of 1250 kg/m$^3$  \citep{Chesley2014}. Thus, it is reasonable to consider that the bulk density of P/2013 R3 may be between 1000 kg/m$^3$ and 1500 kg/m$^3$. The empty triangles and squares in Fig. \ref{Fig:initialPeriod_Cohesion} give the end points for each curve; therefore, the actual configuration could likely be between these points. This provides the following two interpretations.

First, although the spin curve by \cite{Pravec2007} implies that the spins of asteroids ranging from 0.4 km to 1.0 km in size may be bounded at the spin barrier, 2.2 hr, our result suggests that the breakup of P/2013 R3 could have occurred at a shorter spin period than the spin barrier. As shown in Fig. \ref{Fig:initialPeriod_Cohesion}, the slowest spin period is 1.9 hr, occurring when $\rho=$1000 kg/m$^3$, $2 a = 1.0$ km, and $\Delta v = 0.2$ m/s. This condition may be an extreme case in our problem. Again, since the effective radius measured from the October 1, 2013 data is still affected by the dust cloud, \cite{Jewitt2014B} state that the 330 m effective radius measured from the December 13, 2013 data is more accurate. Thus, we believe that a size of 1.0 km ($> 2 \times 330$ m) may be too large. This explains that the initial spin of this asteroid should be faster than the spin barrier. 

Second, the possible cohesive strength ranges from 40 Pa to 210 Pa. This estimate is comparable to that inferred for rubble pile asteroids in \cite{Sanchez2014}. If P/2013 R3 is a rubble pile and the size distribution of its particles extends down to the $\mu$m level, then van der Waals forces may supply the needed cohesive strength for such an asteroid. The cohesive strength of a cohesive self-gravitating aggregate was determined by \cite{Sanchez2014} to be directly related to the average grain size and the Hamaker constant\footnote{The Hamaker constant is directly related to the strength of the cohesive forces between any two bodies whose surfaces are in contact.}. Using a Hamaker constant of $\sim$0.036 N/m, which is consistent with lunar regolith \citep{Perko1996}, for an asteroid with cohesive strength of 40 - 210 Pa and with a friction angle of 37.5$^\circ$, we end up with an average particle size of 1.2 - 6.1 $\mu$m, consistent with the size range of the sample from (25143) Itokawa \citep{Tsuchiyama2011}. Therefore, it is reasonable to believe that a rubble-pile asteroid with hundreds of meters in size could have cohesive strength matching the values calculated through this analysis. Furthermore, the possibility that P/2013 R3 is a monolith is quite low because (i) for the present case cohesive strength is highly bounded and (ii) cohesive strength of a typical rock is at least on the order of 10 MPa  \citep{Lambe1969, Jaeger1976}. 

Based on the observations by \cite{Jewitt2014B}, we showed the possible ranges of the initial spin period and cohesive strength of P/2013 R3. However, once additional observations of this asteroid are carried out, they will provide information that can give further constraints on this breakup event. Also, detailed analysis of the subsequent breakups in this event can also be used to develop additional constraints on these parameters. 


%% Included in this acknowledgments section are examples of the
%% AASTeX hypertext markup commands. Use \url without the optional [HREF]
%% argument when you want to print the url directly in the text. Otherwise,
%% use either \url or \anchor, with the HREF as the first argument and the
%% text to be printed in the second.
%\acknowledgments
The authors wish to thank Dr. David Jewitt, UCLA, for his useful comments on the present work. MH also acknowledges constructive conversation for imaging techniques with Dr. Masateru Ishiguro, Seoul National University and UCLA.




\bibliographystyle{apj}
%\bibliography{mainbelt}  

\clearpage

%% the file, uncomment the \plotone, \plottwo, and
%% If you need a layout that cannot be achieved with \plotone or
%% \plottwo, you can invoke the graphicx package directly with the
%% \includegraphics command or use \plotfiddle. For more information,
%% please see the tutorial on "Using Electronic Art with AASTeX" in the
%% documentation section at the AASTeX Web site,
%% http://www.journals.uchicago.edu/AAS/AASTeX.
%%
%% The examples below also include sample markup for submission of
%% supplemental electronic materials. As always, be sure to check
%% the instructions to authors for the journal you are submitting to
%% for specific submissions guidelines as they vary from

%% This example uses \plotone to include an EPS file scaled to
%% 80% of its natural size with \epsscale. Its caption
%% has been written to indicate that additional figure parts will be
%% available in the electronic journal.

\clearpage

\clearpage

\begin{table}
\begin{center}
\caption{Measured Properties of P/2013 R3}
\label{Table:PhysicalProperty}
\begin{tabular}{l l c}
\hline 
Property & Value & Reference \\
\hline
\hline
Relative Distance, $d$, [km] (Oct. 1, 2014) & 3060 &  \\
Diameter of the Initial Body, $2 a$, [km] & 0.4 - 1.0 & \cite{Jewitt2014B}  \\
Relative Velocity, $\Delta v$, [m/s] & 0.2 - 0.5 & \\
\hline
\end{tabular}
\end{center}
\end{table}

%% The following command ends your manuscript. LaTeX will ignore any text

\end{document}